\documentclass[apjl]{emulateapj}
\usepackage{apjfonts}

\begin{document}

\slugcomment{\bf}
\slugcomment{Accepted by ApJ Letters}

\title{Dynamic Meteorology at the Photosphere of HD 209458b}
\shorttitle{Atmospheric Dynamics of HD 209458b}
\shortauthors{Cooper \& Showman}

\author{Curtis S.\ Cooper\altaffilmark{1}, Adam P. Showman\altaffilmark{1}}

\altaffiltext{1}{Department of Planetary Sciences and Lunar and Planetary
Laboratory, The University of Arizona, 1629 University Blvd., Tucson, AZ 85721 USA; 
curtis@lpl.arizona.edu, showman@lpl.arizona.edu}

\begin{abstract}
\label{Abstract}

We calculate the meteorology of the close-in transiting extrasolar planet HD
209458b using a global, three-dimensional atmospheric circulation model.
Dynamics are driven by perpetual irradiation of one hemisphere of this tidally
locked planet.  The simulation predicts global temperature contrasts of
$\sim\!500$ K at the photosphere and the development of a steady superrotating
jet.  The jet extends from the equator to mid-latitudes and from the top model
layer at 1 mbar down to 10 bars at the base of the heated region.  Wind
velocities near the equator exceed 4 $\rm km\;s^{-1}$ at 300 mbar.  The
hottest regions of the atmosphere are blown downstream from the substellar
point by $\sim\!60\degr$ of longitude.  We predict from these results a factor
of $\sim\!2$ ratio between the maximum and minimum observed radiation from the
planet over a full orbital period, with peak infrared emission preceding the
time of the secondary eclipse by $\sim\!14$ hours.

\end{abstract}

\keywords{planets and satellites: general---planets and satellites: individual
(HD 209458b)---methods: numerical---atmospheric effects}


\section{Introduction}
\label{Introduction}

The transiting planet HD 209458b orbits very closely (0.046 AU) to its parent
star with a period of 3.5257 days \citep{Charbonneau:2000, Henry:2000}.  From
transit depth measurements, the mass and radius of HD 209458b are known fairly
accurately: $0.69 \pm 0.05\,M_{\rm Jupiter}$ and $1.32 \pm 0.05\,R_{\rm
Jupiter}$ \citep{Laughlin:2005}.  The age of the system is estimated to be
$5.2$ Gyr, with uncertainties of $\sim\!10\%$.  Furthermore, owing to
careful measurements of the stellar spectrum during the planet's transit, much
is now known about HD 209458b's atmospheric properties \citep{Brown:2001,
Charbonneau:2002, Vidal-Madjar:2003, Vidal-Madjar:2004}.

Considerable work has been done to model the spectra, physical structure, and
time evolution of extrasolar giant planets (EGPs) \citep{Burrows:2004,
Chabrier:2004, Iro:2005}.  Relatively less effort, however, has been spent on
EGP meteorologies; i.e., global temperature and pressure fluctuations, wind
velocities, and cloud properties.  Preliminary simulations of the circulation
by \citet{Showman:2002} and the shallow-water calculations of \citet{Cho:2003}
suggest that close-in systems like HD 209458b---with strong day-night heating
contrasts and modest rotation rates---occupy a dynamically interesting regime.

In this article, we report on the results from a multi-layer global
atmospheric dynamics model of HD 209458b.  Our results---the existence of a
fast superrotating equatorial jet at the photosphere that blows the hottest
regions downwind---qualitatively agree with previous three-dimensional
numerical simulations by \citet{Showman:2002}.  The simulation presented here,
however, adopts more realistic radiative-equilibrium temperature profiles and
timescales, superior resolution, and a domain that extends deeper into the
interior.  In particular, the simulations of \citet{Showman:2002} could not
accurately predict the day-night temperature difference at the photosphere.
We here predict this temperature difference to be $\sim\!500\,$K, in agreement
with order of magnitude estimates by \citet{Showman:2002}. 

Our calculations have implications for the planet's infrared (IR) light curve.
With the first infrared detections of the transiting EGPs TrES-1 and HD
209458b \citep{Charbonneau:2005, Deming:2005b}, observational constraints on
the meteorologies of close-in giant planets will likely be possible over the
next two years.

\section{Model}
\label{section:Model}

\begin{figure}
\includegraphics[angle=-90, scale=0.75]{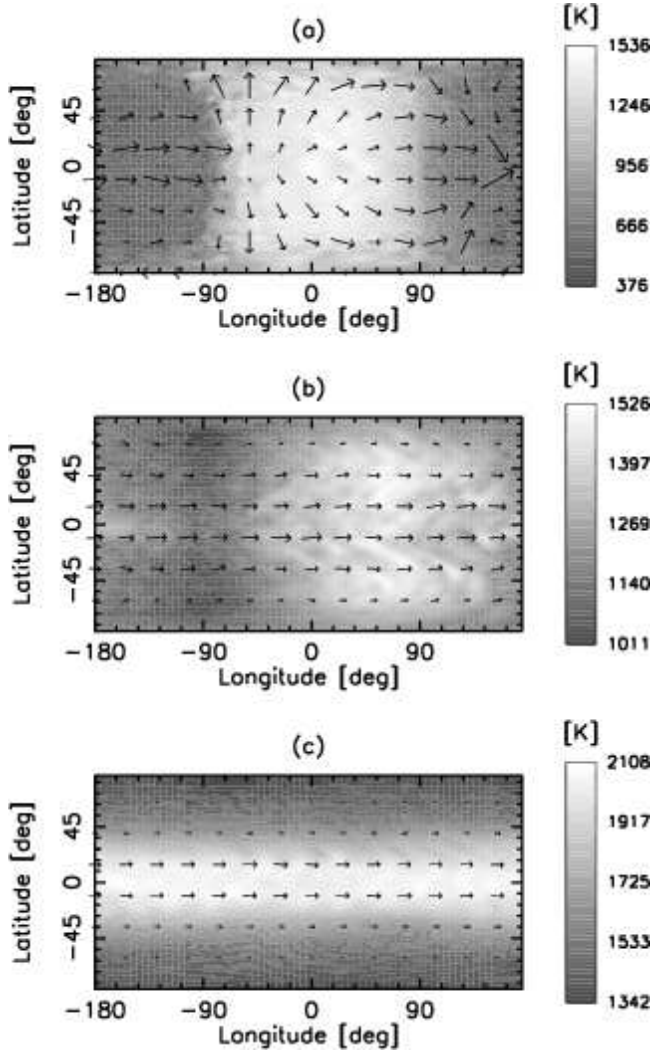}
\caption{Snapshot at 5000 Earth days of simulated temperature (grayscale) and
winds (arrows) on three isobars: 2.5 mbar, 220 mbar, and 19.6 bars in (a),
(b), and (c), respectively.  The substellar point is at $(0,0)$ in longitude
and latitude.  Peak winds are 9.2, 4.1, and 2.8 km sec$^{-1}$ in (a), (b),
and (c).  The simulated temperature difference of $\sim\!500$ K at 220 mbar is
less than the assumed temperature difference of 920 K in radiative-equilibrium
due to advection of hot material from the dayside to the nightside by eastward
winds of $\sim\!\rm 4\;km\;s^{-1}$ on this layer.
}
\label{figure:temperature_layers}
\end{figure}

Our model of the general circulation of HD 209458b integrates the primitive
equations of dynamical meteorology using Version 2 of the ARIES/GEOS Dynamical
Core \citep{Suarez:1995}, which we hereafter abbreviate as AGDC2.  The
primitive equations filter vertically propagating sound waves but retain
horizontally propagating external sound waves (called Lamb waves)
\citep{Kalnay:2003}.  

The grid spacing is $5\degr\times 4\degr$ in longitude and latitude,
respectively ($\sim\!7000$ km near the equator).  We use 40 vertical levels
spaced evenly in log-pressure between 1 mbar and 3 kbar.  This spacing implies
that we resolve each pressure scale height with $\sim\!2$ model layers.  The
scale height ranges from 500--1500 km over the domain of integration. 

Following \citet{Guillot:1996} and \citet{Showman:2002}, we assume synchronous
rotation.  The acceleration of gravity, which does not vary significantly over the
pressure range considered, is set to $g = 9.42\rm\;m\;s^{-2}$.  We take the
mean molecular weight and heat capacity to be constant: $c_p = 1.43\times
10^4\rm\:J\;kg^{-1}\;K^{-1}$ and $\mu = 1.81\times 10^{-3}\rm\;kg\;mol^{-1}$,
which neglects the $\sim\!30\%$ variations in these parameters caused by
dissociation of molecular hydrogen at the deepest pressures in the model.

The intense stellar irradiation extends the radiative zone of HD 209458b all
the way down to $\rm \sim\!1\;kbar$ pressure, with a depth $\sim\!5$--$10\%$
of the planetary radius \citep{Burrows:2003, Chabrier:2004}.  Our integrations
do not solve the equation of radiative transfer directly.  Rather, we treat
the effects of the strong stellar insolation using a Newtonian radiative
scheme in which the thermodynamic heating rate $q$ [W kg$^{-1}$] is given by
\begin{equation}
\label{equation:Cooling}
\frac{q}{c_p} = - \frac{T(\lambda,\phi,p,t) - T_{\rm eq}(\lambda, \phi,p)}{\tau_{\rm rad}(p)},
\end{equation}
which relaxes the model temperature $T$ toward a prescribed
radiative-equilibrium temperature $T_{\rm eq}$.  In Equation
\ref{equation:Cooling}, $\lambda$, $\phi$, $p$, and $t$ are longitude,
latitude, pressure, and time.  The radiative-equilibrium temperature $T_{\rm
eq}$ and the timescale for relaxation to radiative equilibrium $\rm
\tau_{rad}$ are inputs of the AGDC2.  

We rely on the radiative-equilibrium calculations of \citet{Iro:2005} to
specify $T_{\rm eq}(\lambda,\phi,p)$ and $\tau_{\rm rad}(p)$.
\citet{Iro:2005} use the multi-wavelength atmosphere code of
\citet{Goukenleuque:2000} to calculate the radiative-equilibrium temperature
structure of HD 209458b for a single vertical column, hereafter denoted as
$T_{\rm Iro}(p)$.  \citet{Iro:2005} assume globally averaged insolation
conditions (i.e., they redistribute the incident solar flux over the entire
globe).  Their calculation includes the opacities appropriate for a
solar-abundance distribution of gas \citep{Anders:1989}, including the neutral
alkali metals Na and K. 

\citet{Iro:2005} do not consider condensation or the scattering and absorption
of radiation by silicate clouds, which can conceivably form near the
photosphere \citep[see e.g.,][]{Fortney:2003}.  Condensates can potentially
have a significant effect on the planet's radiation balance, depending on the
depth at which they form, the particle sizes, and the vertical extent of cloud
layers \citep{Cooper:2003}.  The net direction of this effect (i.e., to warm
the atmosphere or to cool it) is as yet unclear and remains a subject for
future work.

\citet{Iro:2005} compute radiative-relaxation timescales as a function of $p$
from 0.01 mbar down to 10 bar by applying a Gaussian perturbation to the
radiative-equilibrium temperature profile at each vertical level.  We use
their radiative-relaxation timescales for $\tau_{\rm rad}(p)$ in Equation
\ref{equation:Cooling}.  At pressures exceeding 10 bars, radiative relaxation
is negligible compared to the dynamical timescales considered here.  We simply
assume $q = 0$ on all layers from 10 bar to 3 kbar.

To account for the longitude-latitude dependence of $T_{\rm eq}$ in Equation
\ref{equation:Cooling}, we use a simple prescription.   We choose the
substellar point to be at $(\lambda,\phi) = (0, 0)$. On the dayside, we set
\begin{equation}
\label{equation:T4}
T_{\rm eq}^4(\lambda, \phi, p) = T_{\rm night,eq}^4(p) + [T_{\rm ss,eq}^4(p) - 
T_{\rm night,eq}^4(p)]\cdot \cos(\lambda)\cos(\phi),
\end{equation}
where $T_{\rm ss,eq}(p)$ and $T_{\rm night,eq}(p)$ are the radiative-equilibrium
temperature profiles of the substellar point and night side (assumed to be
uniform over the dark hemisphere), respectively.  Equation \ref{equation:T4}
implies that the hottest $T_{\rm eq}$ profile is at the substellar point.  On
the nightside, we set $T_{\rm eq}(\lambda,\phi,p)$ equal to $T_{\rm
night,eq}(p)$.

We treat the radiative-equilibrium temperature difference between the
substellar point and the nightside, $\Delta T_{\rm eq}(p) = T_{\rm
ss,eq}(p)-T_{\rm night,eq}(p)$, as a free parameter that is a specified
function of pressure.  To determine $T_{\rm ss,eq}(p)$ and $T_{\rm
night,eq}(p)$ from \citet{Iro:2005}'s profile and our specified $\Delta T_{\rm
eq}$, we horizontally average $T_{\rm eq}^4$ on the top layer of our model
over the sphere and set it equal to $T_{\rm Iro}^4$ at that pressure.  Based
on the $\sim\!1000\;$K day-night temperature differences from
\citet{Iro:2005}, we use $\Delta T_{\rm eq} = 1000\rm\;K$ for pressures less
than 100 mb and decrease it logarithmically with pressure down to 530 K at the
base of the heated region (10 bar).  Newtonian cooling is a crude
approximation to the true radiative transfer, but the scheme is
computationally fast---hence allowing extensive explorations of parameter
space---and gives us direct control over the model's diabatic heating.

The model's initial temperature was set to $T_{\rm night,eq}(p)$ everywhere
over the globe; there were no initial winds.  We set the time step equal to
$50\rm\;s$, which is much smaller than the time step required for numerical
stability according to the Courant-Friedrichs-Lewy (CFL) criterion
\citep{Kalnay:2003}.

\section{Results \& Discussion}
\label{section:Simulation_Results}

\begin{figure*}
\includegraphics[scale=0.5]{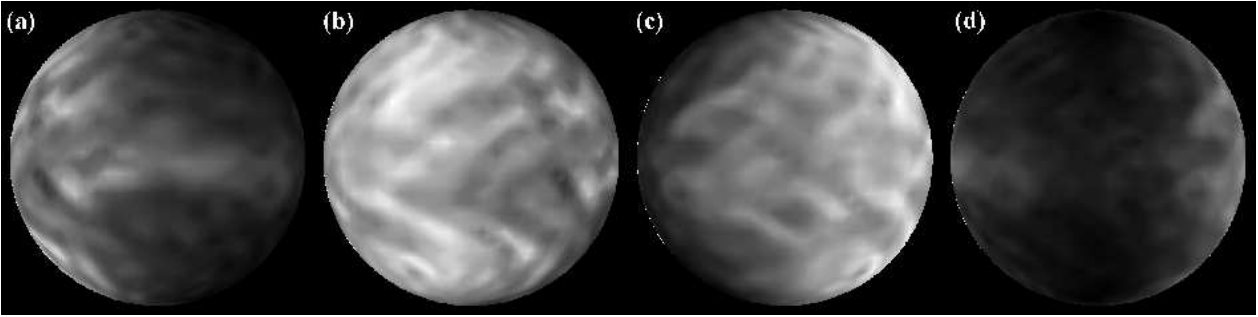}
\caption{Views of the predicted flux emitted from HD 209458b as it would
appear from Earth at four orbital phases: (a) in transit, (b) one-quarter
after the transit, (c) in secondary eclipse, and (d) one-quarter before the
next transit.  The planetary rotation axes are vertical, with the
superrotating jet seen in Figure \ref{figure:temperature_layers} going from
left to right in each panel.  The temperature on this layer ranges from 1011 K
(dark) to 1526 K (bright).  In radiative equilibrium, panel (a) would be
darkest and (c) would be brightest. The globes show that a difference in
observed flux from the planet between the leading and trailing phases of its
orbit---panels (b) and (d)---is a signature of winds.
}
\label{figure:orbital_phases}
\end{figure*}

By 5000 days of simulation time, the simulation has reached a statistical
steady-state, at least down to the 3 bar level, which is the level above which
$99\%$ of the stellar photons are absorbed \citep{Iro:2005}.  Deeper than 3
bars, the kinetic energy continues to increase with time as these layers
respond to the intense irradiation on relatively long timescales $\tau_{\rm
rad}\sim\!\rm 1\;yr$.

At pressures less than 10 bars, the model rapidly develops strong winds and
temperature variability in response to the imposed day-night heating contrast.
The upper atmosphere is nearly in radiative equilibrium (Figure
\ref{figure:temperature_layers}a), with temperature contrasts of
$\sim\!1000\,$K.  This results from the fact that the radiative-equilibrium
time constant $\rm \tau_{rad}$ at 2 mbar is only $\sim\!1$ hour, which is much
shorter than the timescale for winds to advect heat across a hemisphere.  

At 2 mbar pressures, supersonic winds exceeding $9\;\rm km\;s^{-1}$ appear at
high latitudes, with strong north-south as well as east-west flow.  Supersonic
winds are plausible in the dynamical regime of HD 209458b due to the immense
radiative forcing from the parent star.  For comparison, the planet Neptune
has high-velocity zonal jets with wind speeds approaching the local speed of
sound \citep{Limaye:1991}, but the radiative forcing is a million times
weaker than it is for HD 209458b.  

In contrast, the flow near the photosphere near 220 mbar is dominated by an
eastward jet extending from the equator to mid-latitudes (Figure
\ref{figure:temperature_layers}b).  The temperature contrasts reach
$\sim\!500$ K at this level, with the hottest part of the atmosphere advected
$\sim\!60\degr$ downstream from the substellar point by the $\rm
4\;km\;s^{-1}$ eastward jet.  Here, the time constant for relaxation to the
radiative equilibrium $\tau_{\rm rad}$ is several Earth days.  The downstream
advection of the hottest regions results from the fact that the radiative and
advection timescales are comparable at this level.  The small-scale bar-like
features visible in the middle frame of Figure \ref{figure:temperature_layers}
propagate nearly horizontally westward relative to the flow at $\sim\!3\rm
\;km\;s^{-1}$, which is close to the speed of sound.  These features are most
consistent with Lamb waves \citep[][p. 42]{Kalnay:2003}.

At 19.6 bar, the equatorial winds remain extremely
fast---$2.8\rm\;km\;s^{-1}$---but the temperature structure exhibits little
longitudinal variability (Figure \ref{figure:temperature_layers}c).  Our
model's circulation results entirely from radiative heating occuring at
pressures less than $10\;\rm bars$.  Therefore, any winds or temperature
variability that develops at pressures exceeding $10\;\rm bars$ results solely
from downward transport of energy from the overlying heated layers by vertical
advection or wave transport.  The layers between 3--$30\;\rm bars$ contain
$\sim\!70\%$ of the atmosphere's kinetic energy.  Lower pressures have fast
winds but little mass, while the winds drop rapidly to zero at pressures
exceeding $\sim\!30\;\rm bars$.  

The essential assumption of the primitive equations is hydrostatic balance
\citep{Holton:1992, Kalnay:2003}, an approximation generally valid for shallow
flow.  This applies to the case of HD 209458b's radiative zone, which has a
horizontal to vertical aspect ratio of $\sim\!100$.  Analysis of the terms in
the full Navier-Stokes vertical momentum equation reveals that, even in the
supersonic flow regime obtained here, the vertical acceleration and curvature
terms would have magnitudes only $\sim\!1\%$ and $\sim\!10\%$ that of the
hydrostatic terms at the photosphere and top of the model, respectively.  The
caveat to this result is that vertical accelerations can still be important
for sub-grid scale structures.  These simulations also do not include the
possible effects of vertically propagating shocks, which can conceivably
dissipate atmospheric kinetic energy.

To confirm the effects shown in Figure \ref{figure:temperature_layers}, we
have run additional simulations using 1D radiative-equilibrium temperature
profiles from \citet{Burrows:2003} and \citet{Chabrier:2004}, which are
significantly hotter than that of \citet{Iro:2005} due to differing 
assumptions about the heat redistribution.  We have also experimented
with changing the value of the free parameter $\Delta T_{\rm eq}$, which
controls the strength of the radiative forcing in our Newtonian cooling scheme
(Equation \ref{equation:Cooling}).  We have run simulations at $\Delta T_{\rm
eq}$ of 100 K, 250 K, 500 K, 750 K, and 1000 K.  The essential features of the
simulation presented here are representative of the results of the other
simulations performed: they all develop a stable superrotating jet at the
equator extending to the mid-latitudes and a hot region of atmosphere downwind
of the substellar point.  But for a given $\Delta T_{\rm eq}$, the simulated
temperature \em differences \em are not affected by which group's
radiative-equilibrium temperature profile is used.  For values
of $\Delta T_{\rm eq}$ between 500--1000 K, the simulations all produce
temperature contrasts at the photosphere ranging from 300--600 K, with peak
equatorial winds speeds in the range 2--5 $\rm km\;s^{-1}$.  The simulations
employing the atmospheric profiles of \citet{Burrows:2003} and
\citet{Chabrier:2004} do, however, produce hotter (by $\sim\!500\rm \;K$) mean
photospheric temperatures than our nominal case. 

We have also run the model with a different initial condition.  We started
this alternate simulation with identical input parameters to the simulation
shown in Figure \ref{figure:temperature_layers} but with a retrograde zonal
wind profile: $u = {\rm -3\;km\;s^{-1}\;cos^4(\phi)\;tan^{-1}(2\;bar}/p)$.  We
set the meridional wind $v$ to zero and set the pressures to be in
gradient-wind balance with the initial winds \citep{Holton:1992}.  After 5000
days of integration time, a strong superrotating jet at the equator at 220
mbar develops in the simulation, with maximum wind speeds of $\rm 3.4
\;km\;s^{-1}$ and temperature contrasts of $\rm \sim\!430\;K$.  The similarity
with the simulation presented in Figure \ref{figure:temperature_layers} shows
that the overall flow geometry is not strongly sensitive to the initial
conditions.

Our results differ from the one-layer shallow-water simulations of
\citet{Cho:2003}, who find that the mean-equatorial flow for HD 209458b is
westward.  Shallow-water turbulence simulations consistently produce westward
equatorial flow, even for planets such as Jupiter and Saturn whose equatorial
jets are eastward \citep{Cho:1996, Iacono:1999a, Iacono:1999b, Peltier:2001,
Showman:2004}. This effect may result from the exclusion of three-dimensional
momentum-transport processes in one-layer models.  For example,
three-dimensional effects are important in allowing equatorial superrotation
in numerical models of Venus, Earth, and Jupiter \citep{DelGenio:1996,
Saravanan:1993, Williams:2003}.  Nevertheless, the shallow-water turbulence
models successfully capture essential aspects of mid-latitude jets for giant
planets in our solar system \citep{Cho:1996}.  

\section{Orbital Phases and Predicted Lightcurve}
\label{section:Orbital_Phases}

\begin{figure}
\includegraphics[angle=-90, scale=0.5]{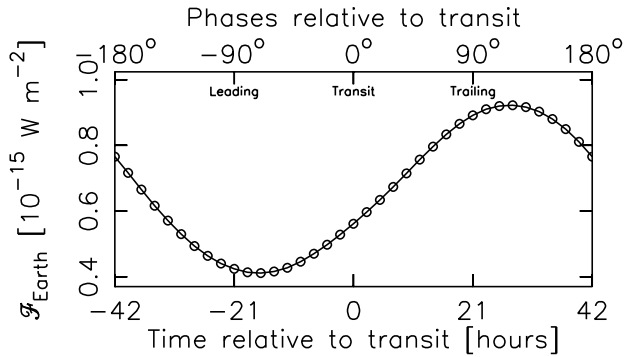}
\caption{Synthetic lightcurve of the 220 mbar layer of our model atmosphere
for HD 209458b.  This simulation predicts peak emission from the planet 14
hours \em before \em the time of the secondary eclipse.
}
\label{figure:light_curve}
\end{figure}

To illustrate the observable effects of the circulation, Figure
\ref{figure:orbital_phases} shows orthographic projections \citep{Snyder:1987}
of the blackbody flux of HD 209458b on the 220-mbar level.  The key feature is
the hot region downstream from the substellar point, which faces Earth after
the transit and before the secondary eclipse.  This pattern differs
drastically from that expected in the absence of winds, in which case the
hottest regions would be at the substellar point.

We show the infrared light curve of the planet as predicted by our simulations
in Figure \ref{figure:light_curve}.  The circles represent the total flux
received at Earth from the planet every $\rm 10\degr$ in its orbit.  The
points are derived by integration of the radiation intensity emitted into a
solid angle projected in the direction toward Earth.  We assume in this
calculation that each column of the atmosphere at the photosphere emits
radiation with the intensity of a blackbody, $\sigma T^4/\pi$, with
temperatures varying as shown in Figure \ref{figure:temperature_layers}(b).
We use 223 mbar for the photospheric pressure, which is the layer closest to
where the effective temperature in \citet{Iro:2005}'s model equals their
computed actual temperature.

The model predicts a phase lead of $60^{\circ}$---or 14 hours---between peak
radiation from the planet and the secondary eclipse, when the illuminated
hemisphere faces Earth.  Based on the $\sim\!500\rm\; K$ temperature contrasts
shown in Figure \ref{figure:light_curve}, we derive a ratio of $2.2$ between
the maximum and minimum flux.  To confirm the validity of this prediction, we
performed analogous light curve calculations assuming the IR flux is emitted
from a single pressure level ranging from 150--450 mbar, where the photosphere
conceivably could be.  The magnitude of the flux ratio ranges from 3.2 to 1.4
over this range of pressures.  Our predicted phase lead of $60\degr$ shifts by
$\pm 20\degr$ over this range of pressures.  These effects are inversely
correlated: if the photosphere pressure is less than 220 mbar, then the flux
ratio increases but the phase shift decreases and vice-versa.  

Nevertheless, the possible formation of clouds high in the atmosphere of HD
209458b ($p < 200\;$mbar)---which we have not treated here---may significantly
alter the radiation budget. Additionally, uncertainties in the input
parameters and the Newtonian cooling approximation limit the precision with
which we can predict the magnitude of the effects described above.  A major
future advance in the characterization of close-in EGP atmospheres will come
from coupling dynamics and radiative transfer directly.  We note, however,
that the inputs to such a model are likely to be somewhat hypothetical as
well, given the limited constraints available from observations.


\acknowledgements

Special thanks to T. Guillot, N. Iro, and B. B\'ezard for valuable guidance on
the radiative time constant prior to publication of their results.  Thanks
also to the referee, J.W. Barnes, J.J. Fortney, P.J. Gierasch, and many others
for advice on the project and manuscript.  Figures created using the free
Python Numarray and PLplot libraries.  This research was supported by NSF
grant AST-0307664 and NASA GSRP NGT5-50462.

\bibliographystyle{apj}

\end{document}